\RequirePackage{ifpdf}
\ifpdf 
\documentclass[pdftex]{sigma}
\else
\documentclass{sigma}
\fi

\numberwithin{equation}{section}

\begin{document}
\allowdisplaybreaks

\renewcommand{\PaperNumber}{022}

\FirstPageHeading

\renewcommand{\thefootnote}{$\star$}

\ShortArticleName{Hidden Symmetries of M-Theory and Its Dynamical
Realization}

\ArticleName{Hidden Symmetries of M-Theory\\ and Its Dynamical
Realization\footnote{This paper is a contribution to the
Proceedings of the Seventh International Conference ``Symmetry in
Nonlinear Mathematical Physics'' (June 24--30, 2007, Kyiv,
Ukraine). The full collection is available at
\href{http://www.emis.de/journals/SIGMA/symmetry2007.html}{http://www.emis.de/journals/SIGMA/symmetry2007.html}}}

\Author{Alexei J. NURMAGAMBETOV}
\AuthorNameForHeading{A.J. Nurmagambetov}

\Address{A.I. Akhiezer Institute for Theoretical Physics, NSC
``Kharkov Institute of Physics\\ and Technology'', 1
Akademicheskaya Str., 61108 Kharkiv, Ukraine}

\Email{\href{mailto:ajn@kipt.kharkov.ua}{ajn@kipt.kharkov.ua}}

\ArticleDates{Received October 31, 2007, in f\/inal form February
06, 2008; Published online February 19, 2008}

\Abstract{We discuss hidden symmetries of M-theory, its feedback
on the construction of the M-theory ef\/fective action, and a
response of the ef\/fective action when locality is preserved. In
particular, the locality of special symmetries of the
duality-symmetric linearized gravity constraints the index
structure of the dual to graviton f\/ield in the same manner as it
is required to separate the levels 0 and 1 generators subalgebra
from the inf\/inite-dimensional hidden symmetry algebra of
gravitational theory. This conclusion fails once matter f\/ields
are taken into account and we give arguments for that. We end up
outlining  current problems and development perspectives.}


\Classification{83E15; 83E50; 53Z05}

\renewcommand{\thefootnote}{\arabic{footnote}}
\setcounter{footnote}{0}

\section{Introduction}

The hidden symmetry structure of M-theory is a subject of
considerable interest during the last decade. It is caused by
lacking the complete dynamics of M-theory with non-perturbative
degrees of freedom, and by our believe that any progress in
understanding the symmetry basis of M-theory is helpful in
searching for the underlying dynamical principle.

{Substantial progress in this direction was recently achieved
within the conjectured, at early stages of the development,
algebraic structure of M-theory. This structure is realized as the
very-extension of the hidden symmetry algebra of dimensionally
reduced ${\rm D}=11$ supergravity~\cite{west01}.} Though many
arguments in favor of the conjecture were subsequently found, this
subject is currently under debates. Nevertheless, seminal ideas
of~\cite{west01} stimulated the development of a new special type
of inf\/inite-dimensional algebras, the so-called very-extended
algebras~\cite{gow02}, that was resulted in recognizing the
special r\^{o}le of Kac--Moody-type algebras in M-theory
setting~\cite{Buyl06}.

Another important consequence arising from the results of
\cite{west01,west02} was the duality-symmetric structure those of
M-theory bosonic tensor f\/ields which cast the bosonic subsector
of \mbox{${\rm D}=11$} supergravity.  Inclusion of their dual
f\/ields is strongly
 expected once we take non-perturbative degrees of freedom into
account. Their realizations may be dif\/ferent, a M5-brane
\cite{blnpst97,apps97} is one of them. Getting of M5-branes
requires a suf\/f\/icient modif\/ication of ${\rm D}=11$
supergravity action \cite{BBS98}, which becomes duality-symmetric
with respect to the third and sixth rank tensor f\/ields. The
corresponding generators can be found on the hidden symmetry
algebra side \cite{west01}.

{With account of the above-mentioned points, the duality-symmetric
${\rm D}=11$ supergravity action \cite{BBS98} can be considered as
a good staring point in searching for the least action principle
of hidden constituents of M-theory which are encoded in the
symmetry algebra.} However, the construction of \cite{BBS98} has
to be suf\/f\/iciently extended with new f\/ields. {They would
correspond} to inf\/initely many generators of the conjectured
very-extended symmetry algebra. Steps on this way are discussed in
what follows in more detail.

At the same time we have to point out that the manifestly
covariant Lagrangian approach to duality-symmetric theories
\cite{pst}, which is in the focus of the paper, is not the only
way to construct the least action principle of M-theory. Other
ways {(see e.g.~\cite{dhn02,eh04})}, which also exploit the
inf\/inite-dimensional structure of the M-theory hidden symmetry
algebra, are subjects of recent reviews \cite{Buyl06,hpp07}. The
alternative least action principle mentioned there is based on a
sigma-model (propagating in one time-like direction, so a
particle-type) action invariant under an inf\/inite-dimensional
algebra transformations. Realization of this program is very
attractive (see~\cite{dn07}), however, some conceptual points
should be recovered on the way. For instance, it is a questionable
point on the consistent coupling of a dynamical M5-brane and other
brane sources to such a sigma-model-type action with retaining the
(special) symmetries of branes. Other points of further
development are {the extension of the `dictionary' between
sigma-model variables and space-time f\/ields beyond low-levels
(see~\cite{dhn02,hpp07} for details) and} the generalization of
the approach to the {completely supersymmetric} case\footnote{It
is worth mentioning that supersymmetry plays an important r\^{o}le
in realising the hidden symmetry structure (see also footnote 6 in
the paper). However, the generalization of the discussed
construction to the supersymmetric case may cause a trouble. It
would require the non-linear realization of super-algebras, which,
in turn, may require the (of\/f-shell) formulation in
superf\/ields. However, the upper limit of the (on-shell)
supergravity superf\/ield formulation does not exceed $N=4$
supersymmetry, while ${\rm D}=11$ case requires $N=8$.}.
Nonetheless, one may notice an apparent advantage of the approach:
the sigma-model-type action is based on the non-linear realization
of the hidden symmetry algebra, hence the feedback of the algebra
structure on the sigma-model dynamics is manifest on this
way\footnote{Let us also mention the sigma-model-type action of
the duality-symmetric ${\rm D}=11$ supergravity \cite{bns04} which
developed ideas of \cite{cjlp98}. The extension of the
construction of \cite{bns04} to include M2 and M5 brane sources,
as well as the construction of the sigma-model-type action for the
duality-symmetric type IIA supergravity were done
in~\cite{ajn04IMP,ajn04Du,ajn04PZh}.}.

The extension of the duality-symmetric ${\rm D}=11$ supergravity
action \cite{BBS98}, in its bosonic subsector, with the graviton
dual f\/ield was proposed in \cite{AJN04}. Such an extension
required introducing non-locality. The non-locality of the
proposed action, the symmetry structure and dynamics were subjects
of intensive discussion in our previous paper \cite{AJNSigma}.

In this paper we extend the analysis of the duality-symmetric
linearized gravity, made in \cite{AJNSigma}, and establish the
restriction on the index structure of the graviton (or vielbein in
the f\/irst order approach) dual f\/ield
\[
A_{[a,b_1 \dots b_{D-3}]}=0,
\]
which ensures the locality of the special symmetries (see
\cite{pst}). Remarkably, the similar constraint, but on the hidden
symmetry algebra side, was found in \cite{west01,west02}. This
constraint separates the subalgebra of generators corresponding to
the graviton and to the graviton dual f\/ield from the rest of the inf\/inite-dimensional algebra, which is the hidden
symmetry algebra of gravitational theory. On our side this
constraint is required for retaining the locality of the model,
and since the vielbein and its dual partner are related to each
other via the duality relation, the constraint on the dual f\/ield
removes the antisymmetric part of the originally unconstrained
vielbein. Put it dif\/ferently, the locality of the linearized
duality-symmetric gravity results in the constraint on the
vielbein which leads to the Fierz--Pauli-type linearized spin-2
theory.

We should warn the reader that the obtained result takes place for
the pure duality-symmetric linearized gravity. Once matter
f\/ields are included, the action of the model becomes non-local
and the of\/f-shell locality cannot be kept anymore.

The organization of the paper is as follows. To f\/ix ideas and to
make the paper self-contained we  brief\/ly review dualities of
String Theory and their connection to the hidden symmetries of
M-theory (Section~\ref{sec2}). Next, we discuss the algebraic
structure of M-theory based on \cite{west01} (Section~\ref{sec3}),
and its restriction to the gravity case (Section~\ref{sec4}). In
Section~\ref{sec5} we discuss the realization of the
duality-symmetric gravity~\cite{AJN04} within the approach
of~\cite{pst}, Section~\ref{sec6} contains extended, in comparison
to~\cite{AJNSigma}, analysis of the linearized theory. To make a
contact to M-theory we consider the duality-symmetric linearized
gravity in presence of matter f\/ields (Section~\ref{sec7}). We
give arguments on the of\/f-shell non-locality in the case, which
is general and do not depend on the nature of matter f\/ields.
Summing up of the results is made in Conclusions. The notation of
the paper can be found in Appendix.


\section{String Theory dualities and hidden symmetries}\label{sec2}

\begin{figure}[h!]
\centering
\includegraphics[width=.86\textwidth]{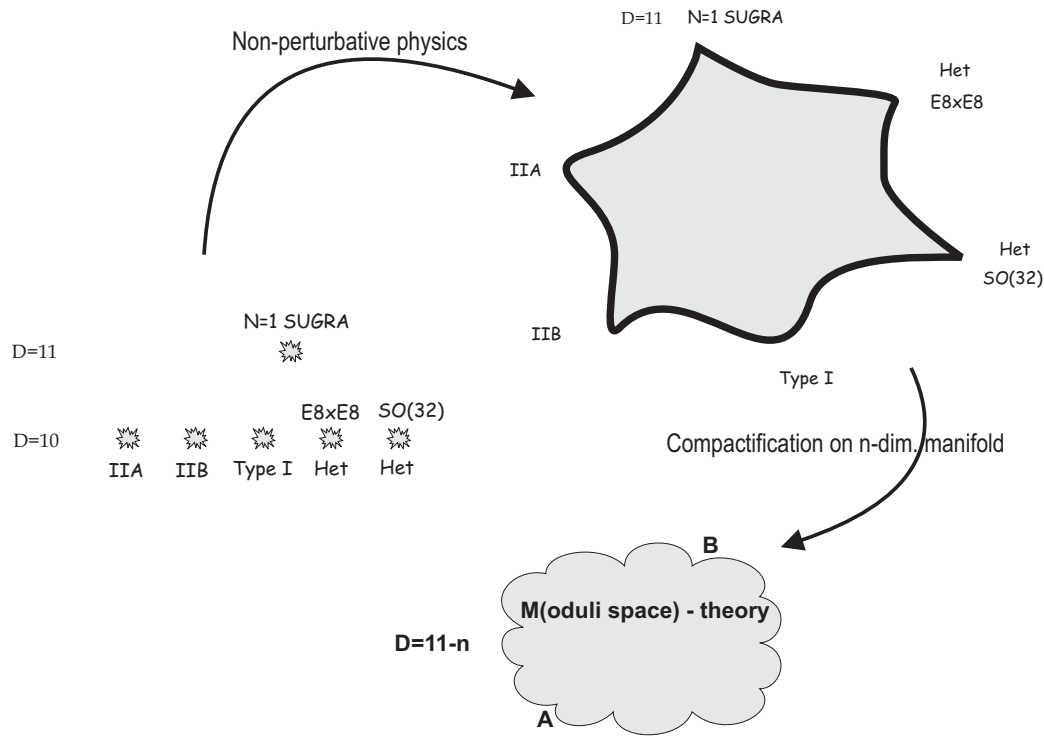}
\caption{The map of String Theory.} \label{figure:fig1}
\end{figure}

Dualities and hidden symmetries of String Theory are closely
related to each other. To realize this relation we begin with the
following cartoon of String Theory (Fig.~\ref{figure:fig1}).

The left part of Fig.~\ref{figure:fig1} is a cartoon of String
Theory within the perturbative framework, with six disjoint points
of f\/ive dif\/ferent superstring theories in ${\rm D}=10$
space-time dimensions and ${\rm D}=11$ $N=1$ supergravity. When
non-perturbative degrees of freedom are taken into account it
results in M-theory description of String Theory with the same six
points, but jointing together. In the bottom of
Fig.~\ref{figure:fig1} one f\/inds an ef\/fective low-dimensional
theory which follows from M-theory after compactifying additional
coordinates. Properties of the ef\/fective theory are essentially
depended on the geometry of internal manifold.

Compactifying M-theory, one arrives at M(oduli space)-theory,
which depends on moduli, i.e.\ some parameters of an ef\/fective
theory arising upon the compactif\/ication. The moduli, but rather
transformations of the moduli under (hidden) symmetry groups, form
the moduli space, dif\/ferent points of which (points A and B on
Fig.~\ref{figure:fig1}) correspond to dif\/ferent ef\/fective
coupling regimes. The ef\/fective coupling, say in the A-point,
may become weak, so one can study the ef\/fective theory
perturbatively there. But~A and~B points of the Moduli space are
related to each other via Duality, and it makes possible to
predict the behavior of theory in the strong coupling point B by
studying the theory in the weak coupling point~A.

\begin{figure}[h!]
\centering
\includegraphics[width=.80\textwidth]{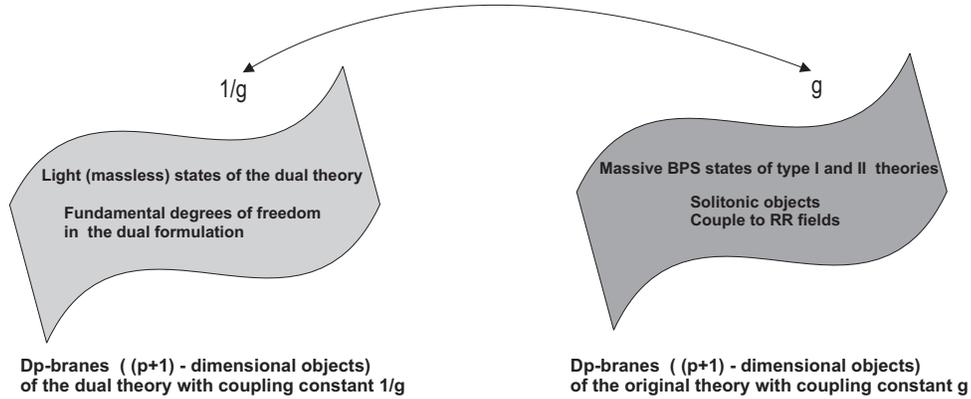}
\caption{{Properties of Dp-branes.}} \label{figure:fig2}
\end{figure}

There are three types of Dualities which connect points in the
Moduli space. Annotating on them we will tightly follow
\cite{Morrison04}. We get started with S-duality, the duality
between strong and weak coupling regimes of the same or
dif\/ferent type theories. It widely applies for analysis of
non-perturbative ef\/fects due to Dp-branes, properties of which
are collected in Fig.~\ref{figure:fig2}.

\begin{figure}[h!]
\centering
\includegraphics[width=.60\textwidth]{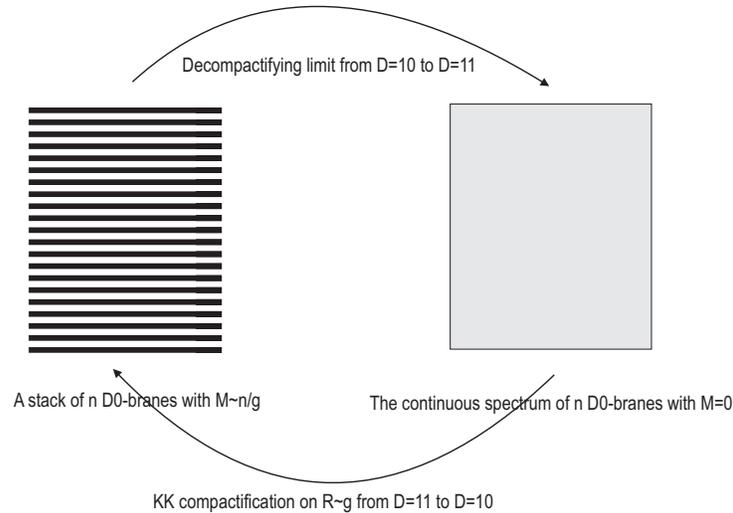}
\caption{S-duality of type IIA String Theory and M-theory.}
\label{figure:fig3}
\end{figure}

In what follows we will focus on maximally supersymmetric, i.e.\
type IIA/IIB String Theories. Type IIB superstring theory is
invariant under S-duality that, together with the invariance under
constant shifts of Ramond--Ramond (RR) f\/ields, results in
$SL(2,R)$ symmetry of the theory (which becomes $SL(2,Z)$ after
the quantization). On the type IIA side S-duality has a
dif\/ferent realization. A stack of $n$  D0-branes with masses
$M\sim n/g$  gets transformed into a smooth spectrum of massless
particles in the strong coupling constant limit $g\rightarrow
\infty$. Such a process may be interpreted as a
decompactif\/ication of type IIA ${\rm D}=10$ string theory into a
${\rm D}=11$ theory (see Fig.~\ref{figure:fig3}).

Once the latter point is accepted, the spectrum of type IIA $n$
D0-branes naturally arises upon the compactif\/ication of ${\rm
D}=11$ theory on the circle of radius $R\sim g$. Hence, the strong
coupling limit of type IIA theory is indeed a theory in ${\rm
D}=11$, referred to as M-theory, and type IIA String Theory is
S-dual to M-theory.

\begin{figure}[h!]
\centering
\includegraphics[width=.25\textwidth]{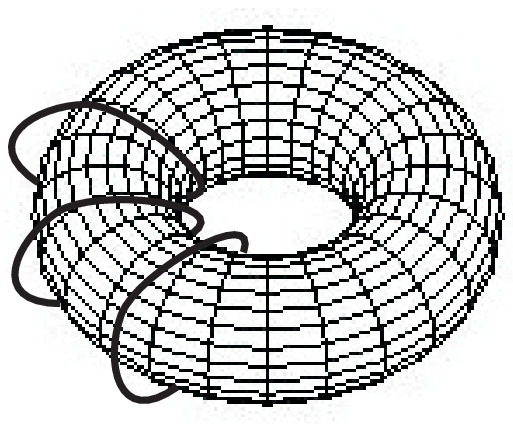}
\caption{Wrapping of a string on an internal torus.}
\label{figure:fig4}
\end{figure}

Another type of duality, Target-space  duality (or T-duality),
arises when a string is embedded into a target space of the
following conf\/iguration ${\cal M}_D={\cal M}_{D-n}\times T^n$,
where $T^n$ is a $n$-dimensional internal torus under which a
string is wrapped $m$ times (see Fig.~\ref{figure:fig4}).

We are interested in the structure of T-duality group, which can
be established as follows. The metric tensor on a $n$-dimensional
torus has the same number of degrees of freedom as that of the
following coset space
\begin{gather*}
\frac{SL(n)}{SO(n)}\times R^+.
\end{gather*}
Here $R^+$ is the torus volume parameter. Since we are dealing
with String Theory, there is also a two-form gauge f\/ield $B_2$,
whose contribution into degrees of freedom on the torus is
$\Lambda^2 R^n=n(n-1)/2$. The total contribution of the metric and
of the two-form gauge f\/ield is matched with the number of
degrees of freedom carried by the following coset space
\begin{gather*}
 \frac{O(n,n)}{O(n)\times O(n)}.
\end{gather*}
The latter is the T-duality group.

What is worth mentioning here is the enhancement of the gravity
internal degrees of freedom global symmetry group, from $SL(n)$ to
$O(n,n)$, due to the contribution of String Theory gauge
f\/ield~$B_2$. We will see in what follows that this statement is
general.

We end up with U-duality, which unites the dualities mentioned in
the above. To establish the U-duality group, one should study both
type IIA/IIB theories in dif\/ferent coupling regimes and in
${\cal M}_{D}={\cal M}_{D-n}\times T^n$ space-times. In type IIA
picture we have $SL(n)$ to  $O(n,n)$ enhancement due to T-duality,
and $SL(n)$ to $SL(n+1)$ enlargement through the M-theory
interpretation (S-duality). These symmetries jointly generate the
larger U-duality group. A convenient way to establish the
U-duality group comes as follows \cite{Morrison04, AJNAST07}.

\begin{figure}[h!]
\centering
\includegraphics[width=.32\textwidth]{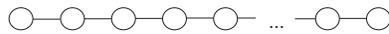}
\caption{$A_{n-1}$ Dynkin diagram.} \label{figure:fig5}
\end{figure}

$SL(n)$ algebra corresponds to $A_{n-1}$ Dynkin diagram with $n-1$
nodes (see Fig.~\ref{figure:fig5}).

\begin{figure}[h!]
\centering
\includegraphics[width=.32\textwidth]{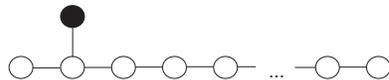}
\caption{$D_n$ Dynkin diagram.} \label{figure:fig6}
\end{figure}

\begin{figure}[h!]
\centering
\includegraphics[width=.36\textwidth]{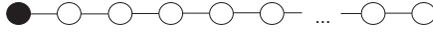}
\caption{$A_n$ Dynkin diagram.} \label{figure:fig7}
\end{figure}

\noindent The enlargement to $O(n,n)$  corresponds to $D_n$
Dynkin diagram with $n$ nodes (Fig.~\ref{figure:fig6}), while the
enlargement to $SL(n+1)$ algebra gets $A_{n}$ diagram (with $n$
nodes, Fig.~\ref{figure:fig7}).

\begin{figure}[h!]
\centering
\includegraphics[width=.36\textwidth]{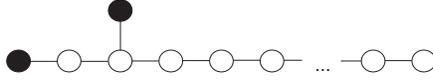}
\caption{$E_{n+1}$ Dynkin diagram.} \label{figure:fig8}
\end{figure}

An entanglement of two diagrams, Fig.~\ref{figure:fig6} and
Fig.~\ref{figure:fig7}, is realized in $E_{n+1}$ diagram,
Fig.~\ref{figure:fig8} (with $n+1$ nodes).

\noindent The latter group is the hidden symmetry global group of
String Theory in ${\cal M}_D={\cal M}_{D-n}\times T^n$
target-space\footnote{One may wonder why the entanglement of
Fig.~\ref{figure:fig6} and Fig.~\ref{figure:fig7} is not realized
in $D_{n+1}$ diagram? The answer is that the reps. of $D_{n+1}$
are not suf\/f\/iciently large to contain massive modes of
strings. Another explanation comes from the above-mentioned
statement on the symmetry enhancement due to String Theory gauge
f\/ield $B_2$. The solid node on the top of Fig.~\ref{figure:fig8}
precisely corresponds to the contribution of this f\/ield (see
e.g.~\cite{lw01,ajn2b}).}.

A symmetry group of String Theory should also incorporate the
symmetry groups of the low-energy ef\/fective actions, viz.\
supergravities. Since ${\cal M}_D={\cal M}_{D-n}\times T^n$ target
space conf\/iguration can be interpreted as the toroidal
reduction, (a subgroup of) $E_{n+1}$ should appear in the
toroidally reduced maximal supergravities. Normally the structure
of $E_{n+1}$ is hidden and is recovered after making additional
steps like, for example, dualisation of f\/ields.

\begin{figure}[h!]
\centering
\includegraphics[width=.55\textwidth]{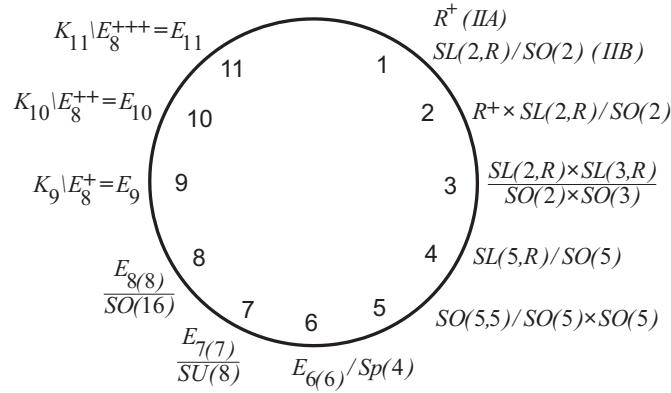}
\caption{M(oduli)-theory clock.} \label{figure:fig9}
\end{figure}

An interpretation of $E_n$  for $n<3$  is subtle (as well as for
high $n$, since $E_8$  is the end of $E_n$ sequence of classical
algebras), rather it is a unifying notation for global symmetry
groups of the moduli space in the reduced theories. Taking into
account the relation of ${\rm D}=11$ $N=1$ supergravity to type
IIA supergravity via the reduction on a one-torus, the $E_n$
sequence of hidden symmetries can be assigned to {the toroidally
reduced} ${\rm D}=11$ supergravity, and should be incorporated
into M-theory. The moduli space of the reduced M-theory in the
low-energy approximation is as in Fig.~\ref{figure:fig9}.

An amazing fact one can read of\/f  Fig.~\ref{figure:fig9} is that
when the reduction goes over three-dimensional space-time down to
dimensions two, one and zero, the $E_n$ sequence of global
symmetry algebras still continues. When $n>8$, the global symmetry
algebras become Kac--Moody-type inf\/inite-dimensional algebras.
It is absolutely unclear why ``conspiracy'' arises, unless it has
presented in the unreduced theory. Following this way, we arrive
at the West's conjecture on $E_{11}$ as a~hidden symmetry algebra
of M-theory~\cite{west01}.

\section[The algebraic structure of $E_{11}$]{The algebraic structure of $\boldsymbol{E_{11}}$}\label{sec3}

The conjecture by West~\cite{west01} is very attractive since we
have nothing hidden to search anymore. In its turn, it has several
non-trivial corollaries. We have noticed that upon the reduction
the symmetry groups get extended from f\/inite-dimensional groups
and algebras to inf\/inite-dimensional Kac--Moody-type algebras.
Hence M-theory constructed this way contains inf\/initely many
massless f\/ields. Some of them may be auxiliary f\/ields, which
do not carry dynamical degrees of freedom. So we have to f\/ind
the relation between f\/ields corresponding to generators of the
Kac--Moody-type algebra and those of perturbative string
spectrum\footnote{The inf\/inite tail of string modes contains in
general massive f\/ield. Therefore, the true matching between
f\/ields of the Kac--Moody-type algebra and string modes can be
made only after f\/iguring out a mechanism of the mass
generation.}.

To make this matching one should know the generators of~$E_{11}$.
Some of them are easily determined from the $E_{11}$ Dynkin
diagram \cite{west01,west02}.

\begin{figure}[h!]
\centerline{\includegraphics[totalheight=0.64in]{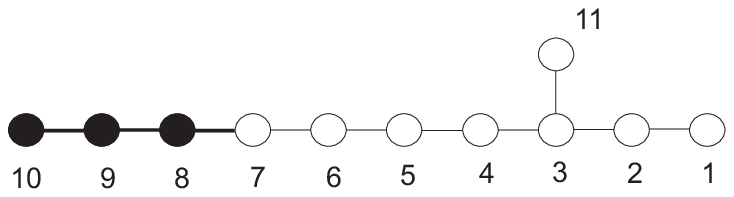}}
\label{Fi:E8+++}
\end{figure}

Deleting the node $11$ results in $A_{10}\sim SL(11)$ algebra. It
corresponds to the gravity sector of ${\rm D}=11$ supergravity
(the so-called gravity line) \cite{west02,lw01}. In such a
decomposition the simple roots of $E_{11}$ are those of $A_{10}$
and $\alpha_{11}=x-\lambda_3$. Here $x$ is orthogonal to the
simple roots of $A_{10}$, and $\lambda_3$ is one of the
fundamental weights of~$A_{10}$.

Any root of $E_{11}$ can be written as a combination of simple
roots $\alpha=\sum\limits_{i=1}^{10}  n_i \alpha_i+l \alpha_{11}$.
The integer $l$ is called the level; it def\/ines the number of
times the simple root $\alpha_{11}$ appears in the root
decomposition. Our choice of the f\/inite-dimensional subalgebra
corresponds to considering the adjoint reps.\ of $E_{11}$ in terms
of the reps.\ of~$A_{10}$.

The following basic facts (see e.g.~\cite{Kac83}) are helpful to
def\/ine the reps. of $E_{11}$ at f\/irst three levels:
$A^{-1}_{ij}=\lambda_i\lambda_j$, $i,j=1,\dots, 10$ (for
simply-laced algebras);
\[
A^{-1}_{jk}=\frac{j(11-k)}{11}, \quad j\ge k; \qquad
A^{-1}_{jk}=A^{-1}_{kj},\quad k\ge j;
\]
$\alpha^2=2,0,-2,-4,\dots$ (for a Kac--Moody algebra with the
symmetric Cartan matrix $A_{ij}$). Summing up the above we get the
following generators at levels $0$, $1$, $2$ and~$3$ \cite{west02}
\begin{gather*}
l=0,~{K^a}_b \quad {\mathrm{of}}\quad A_{10};\qquad
l=1,~R^{[abc]};
\nonumber\\
l=2,~R^{[abcdef]};\qquad l=3,~R^{[abcdefgh],i}.
\end{gather*}
They correspond to graviton, a 3rd rank tensor f\/ield of ${\rm
D}=11$ SUGRA, its 6th rank dual partner, and the dual to graviton
f\/ield. There also is another level 3 generator
$R^{[abcdefghi]}$, which does not occur in $E_{11}$ since the
dimension of a linear space corresponding to this generator (the
so-called multiplicity) is equal to zero.

The obtained generators form the following, non-closed, subalgebra
\cite{west01} which is a part of $E_{11}$
\begin{gather}
[{K^a}_b,{K^c}_d]=\delta^c_b {K^a}_d-\delta^a_d {K^c}_b,\qquad
[{K^a}_b,R^{cde}]=\delta_b^c R^{ade}+\delta_b^d R^{aec}+\delta_b^e
R^{acd}\equiv 3\delta_b^{[c}R^{|a|de]},
\nonumber\\
\label{saE11} [R^{abc},R^{def}]=2R^{abcdef}, \qquad
[R^{abcdef},R^{ghi}]=R^{abcdefgh,i}+R^{abcdefig,h}+R^{abcdefhi,g},
\\
[{K^a}_b,R^{c_1\dots c_6}]=6 \delta_b^{[c_1}R^{|a|c_2\dots
c_6]},\quad [{K^a}_b,R^{c_1\dots c_8,d}]=8
\delta_b^{[c_1}R^{|a|c_2\dots c_8],d}+\delta_b^d R^{c_1\dots
c_8,a}. \nonumber
\end{gather}
One can continue constructing the generators\footnote{By use of,
e.g., \texttt{SimpLie} program (see~\cite{bnb07}) to this end.}
(we recall that there are inf\/initely many generators
of~$E_{11}$), but they will not be in such a transparent
correspondence with f\/ields anymore~\cite{ksw03}. Nevertheless,
we have got enough information to make an intermediate conclusion:
The standard f\/ields of ${\rm D}=11$ SUGRA (the graviton, the 3rd
rank tensor f\/ield) have to be included together with their
duals. Hence, at low-energies, we deal with a {\it
duality-symmetric} formulation of M-theory.

This point is important in context of the ef\/fective dynamical
description of M-theory. It is a well-known fact that the
construction of ${\rm D}=11$ supergravity action with a 6th rank
tensor f\/ield instead of the standard antisymmetric tensor gauge
f\/ield runs into trouble \cite{ntn81,df82}. On the other hand,
the 6th rank tensor f\/ield is needed for coupling of the
dynamical M5-brane \cite{blnpst97,apps97} to ${\rm D}=11$
supergravity. The problem is only overcome within the
duality-symmetric formulation of ${\rm D}=11$ supergravity with
3rd and 6th rank tensor f\/ields~\cite{BBS98}. Therefore, the part
of the algebra~(\ref{saE11}) corresponding to tensor gauge
f\/ields f\/its very well the dynamics of M-theory with
``electric''~M2 and ``magnetic'' M5 branes.

But we still have a question on introducing the gravity into the
game. Though the action of~\cite{BBS98} has included the
Einstein-Hilbert action, it is the incomplete action from the
point of view of~$E_{11}$. It has to be completed with the
graviton dual f\/ield.

Another reason to include the graviton dual f\/ield into M-theory
ef\/fective action was noticed in~\cite{AJN04}. There we pointed
out that the direct way of getting the maximally duality-symmetric
type IIA supergravity action (see \cite{bns04}) through the
reduction of the duality-symmetric ${\rm D}=11$ supergravity
action requires another starting point than the construction
of~\cite{BBS98}. Such a~\mbox{${\rm D}=11$} supergravity
{formulation} should be completely duality-symmetric in the
bosonic f\/ields that requires the dualisation of gravity too.

Before doing anything on this way, {let us make a comment
on~(\ref{saE11}). One may notice that the generator of (\ref{saE11}), corresponding to the dual to
graviton f\/ield, appears as an
interplay between two generators which correspond to M2 and M5
branes of M-theory. The way of $R^{[abcdefgh],i}$ coming seems to
be special, so it is natural to} pose the following question: Does
including the graviton dual f\/ield an {artifact} of M-theory, or
it does appear even in pure gravitational theory?

\section{The hidden symmetry algebra of gravitational theory}\label{sec4}

Long ago it was realized that ${\rm D}=4$ gravity reduced to lower
(${\rm D}=3$, ${\rm D}=2$) dimensions possesses unexpected
symmetries (the $SL(2,R)$s by Ehlers \cite{ehlers} and by
Matzner--Misner \cite{matzmis67}). Their interplay leads to an
inf\/inite-dimensional group (the Geroch group~\cite{geroch71})
which acts on the solutions to the Einstein equation in the
background with two commuting Killing vectors~\cite{maison87}. The
structure of the Geroch group was established~in \cite{Julia82},
where it was shown that the inf\/initesimal form of the Geroch
group corresponds to the af\/f\/ine Kac--Moody algebra
$SL(2,R)^+$.

The Ehlers $SL(2,R)$ is established after dualisation of the
Kaluza--Klein vector to a scalar f\/ield upon the reduction from
${\rm D}=4$ to ${\rm D}=3$. This scalar (the axion) together with
$g_{33}$ component of the metric tensor (the dilaton) form
$SL(2,R)/SO(2)$ coset space. The Matzner--Misner $SL(2,R)$ arises
upon the direct reduction from ${\rm D}=4$ to ${\rm D}=2$ and
corresponds to the global symmetry group of the internal two-torus
(the moduli space group).

On account of the discussed $SL(2,R)^+$ Geroch algebra of the
reduced to ${\rm D}=2$ four-di\-mensional gravity and its
subsequent extension to $SL(2,R)^{++}$ upon the reduction to
\mbox{${\rm D}=1$~\cite{Nicolai92}}\footnote{Actually, this result
is established for D=4 N=1 supergravity. The r\^ole of the local
supersymmetry in forming $SL(2,R)^{++}$ is noteworthy (see
\cite{Nicolai92}) for details).}, one may expect the very-extended
Kac--Moody-type algebra $SL(2,R)^{+++}$ in the end. Following the
West's proposal, this algebra should be the true symmetry algebra
of ${\rm D}=4$ gravity. For ${\rm D}$-dimensional gravity this
algebra becomes $SL(D-2,R)^{+++}$. Since $SL(2,R)^{+++} \sim
A_{1}^{+++}$, the hidden symmetry algebra (for ${\rm D}>3$) is
$A_{D-3}^{+++}$.

As for $E_{11}$ one may f\/igure out the generators of
$A_{D-3}^{+++}$ classifying them w.r.t.\ reps.\ of the gravity
line. It corresponds to $A_{D-1} \sim SL(D)$ in the case (see the
following Dynkin diagram of $A_{D-3}^{+++}$).

\begin{figure}[h!]
\centerline{\includegraphics[totalheight=0.94in]{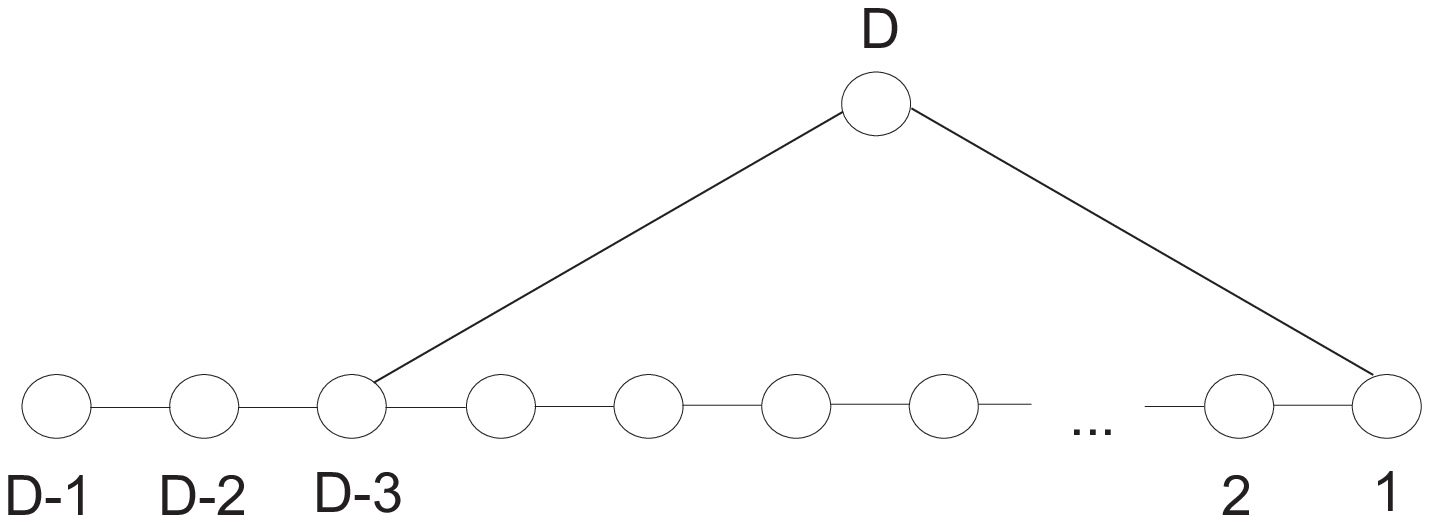}}
\label{Fi:An+++}
\end{figure}

Clearly, we have $A_{D-1}$ generators at level 0 (they are the
generators of $SL(D)$), and the generators $R^{[a_1\dots
a_{D-3}],b}$, $R^{[a_1\dots a_{D-3}b]}$ at $l=1$. It is easy to
recognize the generator corresponding to the dual to graviton
f\/ield, $R^{[a_1\dots a_{D-3}],b}$, while the last generator does
not enter the algebra having the multiplicity zero.

From now on it becomes clear that the presence of the graviton
dual f\/ield is not an artifact of the $E_{11}$ construction. The
corresponding generator enters the symmetry algebra of any theory
which contains gravity. For instance, the following subalgebra of
${\rm D}=11$ pure gravity is included in (\ref{saE11})
\begin{gather}\label{Gr11SA}
[{K^a}_b,{K^c}_d]=\delta^c_b {K^a}_d-\delta^a_d {K^c}_b,\qquad
[{K^a}_b,R^{c_1\dots c_8,d}]=8 \delta_b^{[c_1}R^{|a|c_2\dots
c_8],d}+\delta_b^d R^{c_1\dots c_8,a},
\end{gather}
and this non-closed subalgebra is a part of the
inf\/inite-dimensional algebra $A_{8}^{+++}$. It is also clear
that as soon as the way of constructing the duality-symmetric
gravity will be outlined the generalization to the completely
duality-symmetric ${\rm D}=11$ supergravity will be almost
straightforward.

Before proceeding further, let us make a remark on the index
structure of generators of~(\ref{Gr11SA}). The symmetry properties
of the gravity line generators ${K^a}_b$ are not restricted,
neither to be completely symmetric nor to be completely
antisymmetric. The graviton dual f\/ield generators $R^{a_1\dots
a_{D-3},b}$ are completely antisymmetric over the f\/irst $D-3$
indices. However, if we impose the following additional
restriction on the index structure
\begin{gather}\label{RestGR}
R^{[a_1\dots a_{D-3},b]}=0,
\end{gather}
the subalgebra (\ref{Gr11SA}) splits of\/f the rest of
$A_{8}^{+++}$ and forms its closed part. Equation~(\ref{RestGR})
is also required to close the algebra (\ref{saE11}).

\section{Duality-symmetric formulation of gravity}\label{sec5}

We are turning now to the dynamical realization of the
duality-symmetric M-theory at low-levels of $E_{11}$. It should
include the graviton and its dual f\/ield, so we focus f\/irst on
the construction of the duality-symmetric gravitational theory
action.

The following line of reasoning may be helpful in such a quest.
The standard gravity action
\begin{gather*}
S=\int   d^D x \, \sqrt{|g|} R
\end{gather*}
depends on the dynamical metric tensor f\/ield $g_{mn}=g_{nm}$. To
include its dual f\/ield, we have to add a new term to the action,
which in general depends on the metric tensor, on the graviton
dual f\/ield $\mathcal{G}$ and its `f\/ield strength' ${\mathcal
R}=\partial \mathcal{G}+\cdots$, and possibly on some auxiliary
f\/ields $\xi_i$, which are required for dif\/ferent reasons. It
could be, for instance, the covariantization of the problem.

The duality-symmetric action becomes
\begin{gather}\label{EHadd}
S=\int  d^D x \, \sqrt{|g|} R + \mathcal{L}_{\rm
add.}(g,\mathcal{G},\mathcal{R},\xi_i),
\end{gather}
and its variation over the variables leads to the following set of
equations of motion
\begin{gather}
R_{mn}-\tfrac{1}{2}g_{mn}R+\frac{\delta \mathcal{L}_{\rm
add.}}{\delta g^{mn}}=0, \qquad \frac{\delta \mathcal{L}_{\rm
add.}}{\delta \mathcal{G}}=0, \qquad \frac{\delta \mathcal{L}_{\rm
add.}}{\delta \xi_i}=0.\label{EHaddEOM}
\end{gather}

At f\/irst glance, equations (\ref{EHaddEOM}) describe the
extended, with respect to the original degrees of freedom,
dynamical system. However, we require a special form of the new
term in~(\ref{EHadd}) which, on the one hand, does not spoil the
original dynamics (the set of equations~(\ref{EHaddEOM}) is
reduced to the Einstein equation in the end), and on the other
hand, it should contain some specif\/ic relations, the duality
relations between dual f\/ields, on account of which it could be
possible to recover the original dynamics in terms of the dual
f\/ield.

Another remark concerns the convenient choice of variables to
simplify matching with generators of~(\ref{Gr11SA}). We have noted
that the generators ${K^a}_b$ entering (\ref{Gr11SA}) have the
unrestricted index structure. Therefore, they do not correspond to
the symmetric metric tensor. But they are well f\/itted to
vielbeins~$e^a_m$, whose index structure is also unrestricted.
Moreover, if we treat the upper index of the vielbein as that of
an internal symmetry type, and dualize the vielbein over the lower
index, {similar to} a vector f\/ield, we precisely recover the
dual f\/ield, ${A^{a}}_{[m_1 \dots m_{D-3}]}$, corresponding to
the generator $R^{[a_1 \dots a_{D-3}],b}$. Hence, it is convenient
for our purposes to consider gravity in the f\/irst order
formulation.

Let us make the setting more precise considering the following set
of equations
\begin{gather}\label{EE}
\Sigma_{abc}\cdot R^{bc}+\frac{\delta}{\delta e^a}\left(v\cdot
\mathcal{F}^{a [D-2]}\cdot i_v \mathcal{F}^{b [2]} \eta_{ab}
\right)=0,
\\
\label{AE} d(v\cdot i_v \mathcal{F}^{a[2]})=0,
\\
\label{aE} i_v \mathcal{F}^{a[D-2]} \cdot d(v\cdot i_v
\mathcal{F}_a^{[2]})+i_v \mathcal{F}^{a[2]} \cdot d(v\cdot
\mathcal{F}_a^{[D-2]})=0.
\end{gather}
Written in the dif\/ferential forms notation (see Appendix),
equations (\ref{EE})--(\ref{aE}) are in exact correspondence to
equations (\ref{EHaddEOM}). Equation (\ref{EE}) corresponds to the
Einstein equation extended with contribution of new additional
term, $\mathcal{L}_{\rm add.}$. This term depends on the vielbein
$e^a$ as well as on the vielbein dual f\/ield $A^{a[D-3]}$ through
the following generalized f\/ield strengths
\begin{gather}\label{dr}
\mathcal{F}^{a[2]}= de^a-\ast(dA^{a[D-3]}+\ast\tilde{G}^{a[2]}),
\\
\label{dr1} \mathcal{F}^{a[D-2]}=-\ast\mathcal{F}^{a[2]}.
\end{gather}
The exact def\/inition of $\ast\tilde{G}^{a[2]}$ is not important
for the present discussion, so we skip it for a~while. The rest is
a one-form \cite{pst}
\begin{gather*}
v=\frac{d a(x)}{{\sqrt{-\partial_m a~ g^{mn}~\partial_n a}}},
\end{gather*}
constructed out the auxiliary scalar f\/ield $a(x)$. $\eta_{ab}$
is apparently reserved for the tangent space Minkowski metric
tensor.

One may notice that at least one particular solution to
equations~(\ref{EE})--(\ref{aE}), $\mathcal{F}^{a[2]}=0$, reduces
this system to the single Einstein equation. Moreover, this
particular solution corresponds to one of the f\/irst-order in
derivatives duality relations between f\/ields which ef\/fectively
contain the second order dynamical equations. Taking the external
derivative of $\mathcal{F}^{a[2]}=0$ we obtain the dynamics of
gravity in terms of the dual f\/ield
\begin{gather}\label{Grdual}
d(\ast dA^{a[D-3]})+\dots=0,
\end{gather}
while applying the derivative to the second duality relation,
$\mathcal{F}^{a[D-2]}=0$, results in the following form of the
Einstein equation
\begin{gather}\label{Rem3}
R_{mn}-\tfrac{1}{2}g_{mn}R\equiv d(\ast
de_{a})-{\ast}{\tilde{J}}^{[1]}_{a}=0.
\end{gather}

The exact expression of the ``current'' ${\tilde{J}}^{[1]}_{a}$
has the following, convenient for further references, form
\begin{gather}\label{tilJ}
{\tilde{J}}^{[1]}_{a}=(-)^{\frac{D(D-5)}{2}}{J}^{[1]}_{a}+{\ast}d
{S}^{[D-2]}_{a},
\end{gather}
with
\begin{gather}\label{J}
{J}^{[1]}_{a}={\ast}\big[ {\omega}^{bc}({e})\cdot d{\Sigma}_{{ a}{
b}{ c}}+(-)^{D-3}~{{ \omega}}^{ b}_{~{ d}}({ e})\cdot {\omega}^{{
d}{ c}}({ e})\cdot {\Sigma}_{{ a}{ b}{ c}} \big],
\end{gather}
and
\begin{gather*}
{S}^{[D-2]}_{ a}={\ast} ({e}^{ b}\cdot {e}_{ a}) {e}^{ m}_{ c}
{e}^{ n}_{ b}\partial_{[{m}} {e}^{ c}_{{ n}]}.
\end{gather*}
Note that (\ref{J}) involves the resolved connection
\begin{gather}\label{om}
{\omega}^{{ a}{ b}}({ e})=\tfrac{1}{2} {e}^{c} \big[ {e}^{ m}_{
c}{e}^{{ n}{ a}}\partial_{[{ m}}{e}_{{ n}]}^{ b}-{e}^{ m}_{
c}{e}^{{ n}{ b}}\partial_{[{ m}}{e}_{{ n}]}^{ a}-{e}^{{ n}{
a}}{e}^{{ s}{ b}}
\partial_{[{ n}}{e}_{{ s}]{ c}} \big],
\end{gather}
which follows from the torsion free constraint
\begin{gather}\label{TE}
T^a\equiv de^a-{\omega^a}_b \cdot e^b=0,
\end{gather}
as in the so-called ``one and half'' formalism
(see~\cite{Nieuw81}).

Equations (\ref{EE})--(\ref{aE}) together with equation~(\ref{TE})
can be derived from the f\/irst order action~\cite{AJN04,
AJNSigma} (see Appendix for the notation)
\begin{gather}\label{EHP}
S=\int_{{\mathcal M}^D}\, R^{ab}\cdot \Sigma_{ab}+\tfrac{1}{2}
v\cdot \mathcal{F}^{a [D-2]}\cdot i_v \mathcal{F}^{b [2]}
\eta_{ab}.
\end{gather}
The f\/irst term of (\ref{EHP}) is the standard
Einstein--Hilbert--Palatini action and the second one is a
slightly generalized PST term~\cite{pst}. The special structure of
the latter is important to prove that the duality relation
$\mathcal{F}^{a[2]}=0$ is the general solution to equation
(\ref{AE}) (see~\cite{AJNSigma} in the case), and that the scalar
f\/ield $a(x)$ is the auxiliary f\/ield. Indeed, on the mass-shell
$\mathcal{F}^{a[2]}=0$, the equation of motion of $a(x)$,
equation~(\ref{aE}), is identically satisf\/ied, and does not
carry any additional dynamical information.

Finally, $\tilde{G}^{a[2]}$ entering (\ref{dr}) is def\/ined by
$d\ast \tilde{G}^{a[2]}=\ast\tilde{J}^{a[1]}$. The presence of
$\tilde{G}^{a[2]}$ sacrif\/ices the locality of the action, and of
the generalized f\/ield strengths (\ref{dr}), (\ref{dr1}).
However, the dual f\/ield equation of motion (\ref{Grdual}) can be
rewritten in the local form after the dual f\/ield
redef\/inition~\cite{AJNSigma}.

\section{Duality-symmetric gravity in the linearized
approximation}\label{sec6}

There are numerous indications that the linearized gravity admits
dualisation in the local form (see e.g.
\cite{gopr98,gor98,nieto99,hull98,hull00,hull01,bch03,ht05,jlr05,ahs04}),
and solely in terms of the dual f\/ield. However, the
generalization of the construction to the non-linear case should
be resulted in a duality-symmetric theory. The reasons for that
are as follows.

As for bosonic f\/ields, the f\/ield dual to graviton has to be
described by a second order, in space-time derivatives, equation
of motion. Its structure in a curved background is
\[
 \Box ({\mathrm{ dual~ f\/ield}})+\dots=0,
 \]
where $\dots$ corresponds to possible self-interactions and
interactions with the background gravity f\/ield. The box $\Box$
is the d'Alembertian operator, and this operator is constructed
out the space-time derivatives and the background metric. The dual
f\/ield dynamics within the full non-linear self-consistent theory
(which takes into account the backreaction of the graviton
dynamics)  will contain the d'Alembertian with the dynamical
(non-background) metric. Therefore, the resulted theory will be a
duality-symmetric theory which manages the dynamics of both
f\/ields, the graviton and its dual partner.

Our previous consideration \cite{AJNSigma} of the
duality-symmetric gravity linearization has indicated that
on-shell we encounter the local formulation. Here we would like to
extend the analysis to check the locality of the linearized
duality-symmetric gravity of\/f-shell.

Let us make a quick recap of the on-shell linearized formulation.
We expand the vielbein near the f\/lat space
\begin{gather}\label{linv}
{e}^{ a}(x)=dX^{ m} u^{ a}_{ m}+{\mathcal{E}}^{ a}(x),
\end{gather}
with a constant matrix $u^{ a}_{ m}$. The spin connection
(\ref{om}) linear in $\mathcal{E}^a (x)$ then becomes
\begin{gather}\label{linom}
{\omega}^{{a}{b}}({e}) \cong \tfrac{1}{2}  dX^{ k} u^{ c}_{ k}
 \big[ u^{ m}_{ c} u^{{n}{a}}\partial_{[{m}}{\mathcal{E}}^{
b}_{{n}]} -u^{ m}_{ c} u^{{n}{b}}\partial_{[{m}}{\mathcal{E}}^{
a}_{{n}]}-u^{{n}{a}} u^{{s}{b}}
\partial_{[{n}} {\mathcal{E}}_{{s}] {c}} \big]  +
\mathcal{O}({\mathcal{E}}^2),
\end{gather}
that means
\begin{gather}\label{linJ}
{J}^{[1]}_{ a}\cong \mathcal{O} ({\mathcal{E}}^2).
\end{gather}
Hence the latter expression does not enter the linearized equation
of motion. Furthermore,
\begin{gather*}
{S}^{[D-2]}_{ a}={\ast} \left( dX^{ k} u^{ b}_{ k} \cdot dX^{ l}
u_{{l}{a}} \right) u^{ m}_{ c} u^{ n}_{ b}
\partial_{[{m}} {\mathcal{E}}^{ c}_{{n}]} +
\mathcal{O} ({\mathcal{E}}^2)  \equiv {\mathbb{S}}^{[D-2]}_{
a}+\mathcal{O} ({\mathcal{E}}^2).
\end{gather*}
In ef\/fect, equation of motion (\ref{Rem3}) becomes
\begin{gather}\label{lineom}
d({\ast} d {\mathcal{E}}_{ a} - {\mathbb{S}}^{[D-2]}_{ a})=0,
\end{gather}
and it has the structure of the Bianchi identity $d(\cdots)=0$ of
the dual f\/ield strength. This expression is apparently local, so
the non-locality corresponding to a self-interacting part of the
non-linear action disappears in the linearized limit.

So far we discussed the linearized limit on-shell. Having the
on-shell locality does not guarantee the locality of\/f-shell. Let
us check the locality of the linearized action and the action
symmetries.

The covariant action for the duality-symmetric gravity in the
linearized limit is
\begin{gather}\label{EHPlin}
S=\int_{{\mathcal M}^D}  \left( [R^{ab}\cdot \Sigma_{ab}]_{\tt
{lin.}} +\tfrac{1}{2} v\cdot \mathcal{F}^{m [D-2]}\cdot i_v
\mathcal{F}^{n [2]} \eta_{mn} \right).
\end{gather}
The f\/irst term of (\ref{EHPlin}) is the linearized
Einstein--Hilbert--Palatini action and the second term is the
linearized version of the PST term of~(\ref{EHP}).

The generalized f\/ield strengths which enter~(\ref{EHPlin}) are
def\/ined by
\begin{gather}\label{Flin}
\mathcal{F}_m^{[D-2]}\equiv \mathcal{F}_a^{[D-2]} u_m^a=\left(
dA_a^{[D-3]}-(\ast d\mathcal{E}_a-{\mathbb{S}}_a^{[D-2]}) \right)
u^a_m,  \qquad \mathcal{F}_m^{[2]}=-\ast \mathcal{F}_m^{[D-2]},
\end{gather}
and there is no dif\/ference between f\/lat and curved indices in
the limit. Clearly, the linearized action (\ref{EHPlin}) does not
contain non-local constituents, hence we encounter the locality at
the level of action. But what about the action symmetries?

The covariance of the model is guaranteed by the PST-like
construction of the action, and the local Lorentz transformations,
which act on the tangent f\/lat space indices, also become {truly}
local (see \cite{AJNSigma} for details). The remainder is the
special symmetries of the approach, the so-called PST symmetries
\cite{pst}, to the analysis of which we are turning now.

Varying the action results in
\begin{gather}
\delta S= \int_{{\mathcal M}^D}
 \left(\delta A^{{ m}[D-3]}+\frac{\delta
a}{{\sqrt{-(\partial a)^2}}} i_{v} {\mathcal F}^{{ m}[D-2]}\right)
\eta_{{ m}{n}} \cdot d({v}\cdot i_{ v}{\mathcal
F}^{{ n}[2]}) \nonumber\\
\phantom{\delta S=}{} +\int_{{\mathcal M}^D}   \left(\delta
\mathcal{E}^{ m} +\frac{\delta a}{{\sqrt{-(\partial a)^2}}} i_{ v}
{\mathcal F}^{{ m}[2]}\right) \eta_{{m}{n}} \cdot d({v}\cdot i_{
v}{\mathcal
F}^{{n}[D-2]}) \nonumber\\
\label{varEHPlin} \phantom{\delta S=}{} -\int_{{\mathcal M}^D}
\delta {\mathbb{S}}^{{m}[D-2]} \eta_{{ m}{n}} \cdot {v} \cdot i_{
v} {\mathcal F}^{{n}[2]}  .
\end{gather}
The f\/irst two terms of (\ref{varEHPlin}) vanish under the
following transformations of f\/ields
\begin{gather}\label{PST1}
\delta a({x})=0,\qquad \delta \mathcal{E}^{m}=da\cdot
{\varphi}^{{m}[0]}, \qquad  \delta A^{{m}[D-3]}=da\cdot
{\varphi}^{{m}[D-4]},
\\
\label{PST2} \delta a({x})=\Phi ({x}),\qquad\! \delta
\mathcal{E}^{m}=-\frac{\Phi}{{\sqrt{-(\partial a)^2}}} i_{ v}
{\mathcal F}^{{m}[2]}, \qquad\! \delta
A^{{m}[D-3]}=-\frac{\Phi}{{\sqrt{-(\partial a)^2}}} i_{ v}
{\mathcal F}^{{m}[D-2]}\!\!\!\!\!
\end{gather}
with local gauge parameters $\varphi^{m[0]}$,
{$\varphi^{m[D-4]}$}, $\Phi$. However, the third term of
(\ref{varEHPlin}) does not generally vanish under (\ref{PST1}),
(\ref{PST2}).

Indeed, the variation of this term under (\ref{PST1}) results in
\begin{gather*}
\delta {\mathbb{S}}^{{m}[D-2]} \eta_{{ m}{n}} \cdot {v} \cdot i_{
v} {\mathcal F}^{{n}[2]} \sim v_s \partial_k \varphi^{s[0]} (v^m
{\mathcal{F}_m}^{kl}v_l- v^k {\mathcal{F}_m}^{ml}v_l)-\partial_s
\varphi^{s[0]} {\mathcal{F}_m}^{ml}v_l.
\end{gather*}
One may notice that this variation is equal to zero once
\begin{gather*}
u^a_m \mathcal{F}_{a,[np]}=u^a_m \mathcal{F}_{[a,np]}.
\end{gather*}
In its turn, (\ref{Flin}) requires
\begin{gather*}
u^a_m \mathcal{F}_{a,[n_1\dots n_{D-2}]}=u^a_m \mathcal{F}_{[a,n_1
\dots n_{D-2}]},
\end{gather*}
that leads to
\begin{gather}\label{iff}
u^a_m A_{[a, n_1 \dots n_{D-3}]}=0.
\end{gather}
The invariance of the action under the second special symmetry
(\ref{PST2}) also imposes the constraint~(\ref{iff}). Once this
constraint is relaxed, the PST transformations of the dual f\/ield
$A^{m[D-3]}$ receive non-local corrections.

The encountered constraint on the dual f\/ield index structure
corresponds, on the hidden symmetry algebra side, to
equation~(\ref{RestGR}). Recall, the latter is required to close
the subalgebra~(\ref{Gr11SA}).

\section[Linearized gravity with matter fields and its dualisation]{Linearized gravity with matter f\/ields and its dualisation}\label{sec7}

Having discussed the locality of the duality-symmetric linearized
gravity action based on the algebra (\ref{Gr11SA}), let us take a
further step towards the dynamical realization of the M-theory
algebra~(\ref{saE11}).

To make a contact to M-theory we have to extend the action
(\ref{EHPlin}) at least with a three-form f\/ield kinetic
term\footnote{The non-linear duality-symmetric action of M-theory
based on (\ref{saE11}) can be found in \cite{AJN04,AJNSigma}.}.
Once this gauge f\/ield is taken into account, the expansion of
the vielbein (cf.~(\ref{linv})) gets modif\/ied to the following
form
\begin{gather}\label{S7linv}
 e^a(x)=\tilde{e}^a(x)+\mathcal{E}^a(x),
\end{gather}
where $\tilde{e}^a(x)$ is the solution to the linearized Einstein
equation with the 3-form f\/ield energy-momentum tensor. Putting
it dif\/ferently, we have to expand over the {\it curve}
background in the case rather than over the f\/lat space-time as
it has been done before.

The dif\/ference between a curve space and the Minkowski f\/lat
space becomes clear once we present (\ref{om}) as
\begin{gather*}
\omega^{ab}(e)=\tfrac{1}{2} e^c {\Omega_c}^{ab}(e,\partial e),
\end{gather*}
so expanding the vielbein as in (\ref{S7linv}) we schematically
get
\begin{gather*}
\omega^{ab}(e) \cong \tfrac{1}{2} \tilde{e}^c
{\Omega_c}^{ab}(\tilde{e},\partial \mathcal{E})+\tfrac{1}{2}
\mathcal{E}^c {\Omega_c}^{ab}(\tilde{e},\partial
\tilde{e})+\mathcal{O}(\mathcal{E}^2)+\cdots.
\end{gather*}
Clearly, the latter expansion contains the part linear in the bare
$\mathcal{E}^a$ which was absent in the pure gravity case
(see~(\ref{linom})). As a result, (\ref{linJ}) gets modif\/ied
with terms linear in $\mathcal{E}^a$, and equation~(\ref{lineom})
becomes (for ${\rm D}=11$)
\begin{gather}\label{S7lineom}
d(\ast d\mathcal{E}_a-{S}^{[9]}_{ a})=-\ast J_a^{[1]}.
\end{gather}
The r.h.s.\ of (\ref{S7lineom}) cannot be transformed into a local
curl. Put it dif\/ferently, dualisation of the linearized gravity
with matter requires introducing non-locality. The ``current''
form $\ast \tilde{J}_a^{[1]}$ of~(\ref{tilJ}) is closed but not
exact in the case, so its ``pre-current'' form $\ast
\tilde{G}^{[2]}_a$ ($d \ast \tilde{G}^{[2]}_a=\ast
\tilde{J}_a^{[1]}$) is a non-local expression. Since the
``pre-current'' enters the duality relations (cf.~(\ref{dr}),
(\ref{dr1})) its non-locality induces the non-locality of the
action. Then, symmetries of the action also become non-local, and
(\ref{iff}) does not save the locality.

To make our consideration less formal let us reformulate things in
more familiar fashion. We will do that for ${\rm D}=4$ linearized
gravity with matter; the generalization to a higher-dimensional
case is straightforward. The pure linearized gravity action is as
follows
\begin{gather}\label{h4}
A=\int \, d^4 x \, \partial_a h_{bc}\mathcal{M}^{abcijk}
\partial_i h_{jk}.
\end{gather}
Here $h_{ab}=h_{ba}$ is the linearized graviton f\/ield and
\begin{gather}\label{Mh4}
\mathcal{M}^{abcijk}=(\eta^{ai}\eta^{bc}\eta^{jk}-\eta^{ai}\eta^{bj}\eta^{ci}+2\eta^{ak}\eta^{bj}\eta^{ci}
-2\eta^{ak}\eta^{bc}\eta^{ij})_{\rm symm.}.
\end{gather}
We use the notation of \cite{padma04}; the symmetry properties of
(\ref{Mh4}) is apparent from (\ref{h4}).

Equation of motion which follows from (\ref{h4}) can be written in
the following form
\begin{gather}\label{h4eom}
\tilde{\mathcal{M}}^{abcijk}\partial_a \partial_i h_{jk}=0,
\end{gather}
with $\tilde{
\mathcal{M}}^{abcijk}=\eta^{ai}\eta^{bc}\eta^{jk}-\eta^{ai}\eta^{bj}\eta^{ci}+\eta^{ak}\eta^{bj}\eta^{ci}
-\eta^{ak}\eta^{bc}\eta^{ij}-\eta^{ic}\eta^{jk}\eta^{ab}+\eta^{ak}\eta^{cj}\eta^{bi}$.
Equa\-tion~(\ref{h4eom}) is the Bianchi identity of the dual to
graviton f\/ield, ${U_{rs}}^{|bc}$, which is def\/ined
by{\samepage
\begin{gather}\label{dualh4}
\epsilon^{aprs}\partial_p
{U_{rs}}^{|bc}=\tilde{\mathcal{M}}^{abcijk}\partial_i h_{jk}.
\end{gather}
So, the dualisation is straightforward and does not violate the
locality.}

When the matter source is taken into account, equation
(\ref{h4eom}) gets transformed into
\begin{gather}\label{h4meom}
\tilde{\mathcal{M}}^{abcijk}\partial_a \partial_i h_{jk}=T^{bc}.
\end{gather}
Here $T^{bc}$ is the energy-momentum tensor of matter f\/ields.
Its structure (see, e.g., \cite{padma04,leclerc06} for details) is
as follows
\begin{gather*}
T^{bc}\equiv \left[ \partial^c \phi_A \left( \frac{\partial
L}{\partial(\partial_b \phi_A)}\right)-\eta^{bc} L
\right]+\partial_a \psi^{abc},
\end{gather*}
where $\psi^{abc}=-\psi^{bac}$ is an arbitrary third rank tensor.
Then, equation~(\ref{h4meom}) becomes
\begin{gather}\label{h4meom1}
\partial_a \left( \tilde{\mathcal{M}}^{abcijk} \partial_i h_{jk}-\psi^{abc}\right)=
\left[ \partial^c \phi_A \left( \frac{\partial
L}{\partial(\partial_b \phi_A)}\right)-\eta^{bc} L \right],
\end{gather}
which is nothing but equation~(\ref{S7lineom}) written in terms of
other variables. For a general Lagrangian $L(\phi_A,\partial_a
\phi_A)$ it is quite unexpectable that the r.h.s.\ of
(\ref{h4meom1}) can be presented as a local curl.

Moreover, it cannot be done anyway. If it were done it would be
possible, due to a 3rd rank tensor $\psi^{abc}$, to `neutralize'
the contribution of matter f\/ields to the energy-momentum tensor.
Thus, the energy-momentum tensor could always be set to zero, and
the dynamics of gravity with matter f\/ields would be the same as
the dynamics of pure gravitational f\/ield. Presumably, the latter
is wrong that completes our arguments.

Therefore, the dualisation of the linearized gravity with matter
f\/ields cannot be generally done in the local form. It does not
depend on the nature of matter f\/ields $\phi_A$ which would be
scalar, spinor, vector, tensor or spin-tensor f\/ields.

\section{Conclusions}\label{sec8}

We conclude with the following points. Dualities and hidden
symmetries of String Theory are closely related to each other. It
has been shown that Dualities of String Theory require the
modif\/ication of String Theory to M-theory. The algebraic
structure of M-theory is encoded in hidden symmetries of the
String Theory low-energy ef\/fective actions. Such an algebraic
structure may be realized dynamically in dif\/ferent ways. Here we
have followed \cite{pst, BBS98,AJN04,AJNSigma}.

An essential feature of \cite{AJN04,AJNSigma} is non-locality of
the non-linear action and of the symmetries of the approach. Here
we have focused on the locality of the linearized
duality-symmetric gravity action\footnote{The on-shell locality of
the linearized gravity in dual variables is directly observed from
(\ref{lineom}), (\ref{h4eom}), (\ref{dualh4}).}. We have observed
that the requirement of locality of the linearized
duality-symmetric gravity leads to the constraint on the index
structure of the dual f\/ield (see (\ref{iff})). The corresponding
constraint was previously found on the hidden symmetry of M-theory
\cite{west01,west02} side. Note that this constraint has nothing
to do with the action of the model (\ref{EHPlin}) which is local,
as well as with the equations of motion which follow from the
local action. We have observed that this constraint ensures the
locality of the special gauge transformations,
equations~(\ref{PST1}), (\ref{PST2}), of the linearized
duality-symmetric gravity action. It points at an interesting and
quite unexpected relation between the dynamical local symmetries
of the action and properties of the hidden symmetry algebra.
Furthermore, since the vielbein and the dual f\/ield are related
to each other through the duality relations (\ref{Flin}), the
constraint (\ref{iff}) eliminates the antisymmetric component of
the vielbein ${\mathcal{E}_{[m}}^{n]}$, hence leading to the
Fierz--Pauli-type description of the spin-2 duality-symmetric
linearized theory.

{The encountered constraint, on the hidden symmetry algebra side,
is responsible for closing the subalgebra of the graviton and of
the dual f\/ield generators. It leaves no room for other f\/ields
corresponding to the Kac--Moody-type algebra, inclusion of which
would be helpful, for instance, in quantum description of the
model. Recall that the amplitude of the spin-2 single-particle
exchange violates the Froissart bound, breaking Unitarity at high
energies. Unitarity can be restored but within the Regge poles
theory, where the single-particle exchange is replaced with the
Reggeon exchange, with a bunch of inf\/initely many particles
belonging to the Regge trajectory. It is natural to query whether
the f\/ields of the Kac--Moody algebra has a similar
correspondence to the spin-2 f\/ield as in the Regge theory, but
an answer is unclear for a while.

On the other hand, when the constraint is imposed, we have just a
well-def\/ined, from the point of view of the locality,
duality-symmetric theory, symmetries of which are the standard for
the approach and are well-def\/ined too. If one would try to make
the embedding of the symmetries into an extended symmetry
structure, one should introduce compensator f\/ields, which would
`un-Higgs' the system, thus restoring the large symmetry.
Following this way, we could notice that indeed some of the
compensators would be the true Higgs f\/ields responsible for the
generation of masses of other compensators.}

{Progress in these directions could be achieved on the way of
establishing the correspondence between string theory higher spin
modes and those of the appropriate Kac--Moody algebra.} An
interesting problem {solution to which may help on this way} is to
relax the constraint on the dual to the vielbein f\/ield, and to
include other f\/ields of the Kac--Moody hidden symmetry algebra
into the construction of the duality-symmetric action.

{Finally, let us make a short remark on the sigma-model approach
of \cite{dhn02,eh04,hpp07,dn07}. This approach is based on the
conjecture which claims that the full geometrical data of M-theory
(and ${\rm D}=11$ supergravity as its low-energy limit as well)
can be mapped onto a geodesic motion in the $E_{10}/K(E_{10})$
coset space. The established there `dictionary' between parameters
of the coset space and M-theory f\/ields works good up to the
third level of $E_{10}$ decomposition with respect to $SL(10)$
f\/inite subalgebra. At the third level, where the dual to
graviton f\/ield appears, there is a mismatch between the dynamics
of the coset space parameters and that of ${\rm D}=11$
supergravity bosonic f\/ields. Such a discrepancy may be resolved
with taking into account higher levels of $E_{10}$ on both sides
of the correspondence or with taking into account a~`gradient'
conjecture of~\cite{dhn02}. The latter is equivalent to
introducing the (spatial) non-locality into the theory. We have
encountered the non-locality in the duality-symmetric theory of
gravity with matter f\/ields, so the question is how to realize
the `gradient' conjecture on our side.}

\appendix
\section{Notation and conventions}

Our choice of the signature is the mostly minus. Letters from the
middle of the Latin alphabet are reserved for the curved indices,
letters from the beginning are used for the indices in tangent
space. The Levi-Civita tensor $\epsilon^{a_1 \cdots a_D}$ is
def\/ined by
\begin{gather*}
\epsilon^{01 \dots (D-1)}=1,\qquad \epsilon_{01 \dots
(D-1)}=(-)^{D-1},
\end{gather*}
that implies
\begin{gather*}
\epsilon^{a_1 \dots a_D} \epsilon_{a_1 \dots a_D}=(-)^{D-1} D!.
\end{gather*}

An $n$-form has the following coordinate representation
\begin{gather*}
\omega^{[n]}=\frac{1}{n!} dx^{m_n} \cdots  dx^{m_1}
\omega^{[n]}_{m_1 \dots m_n},
\end{gather*}
and the exterior derivative acts from the right.

The Hodge star is def\/ined by{\samepage
\begin{gather*}
\ast\big(dx^{k_n}  \cdots dx^{k_1}\big)=\frac{1}{(D-n)!}
\frac{\alpha_n}{\sqrt{|g|}}\, dx^{m_{D-n}}  \cdots  dx^{m_1}
{\epsilon_{m_1 \dots m_{D-n}}}^{k_1 \dots k_{D-n}},
\end{gather*}
with coef\/f\/icients $\alpha_n$ f\/ixed to provide the universal
identity
$\ast^2=1$.}

The curvature of $SO(1,D-1)$ connection $\omega^{ab}$ is
$R^{ab}=d\omega^{ab}-{\omega^a}_{c}\cdot \omega^{cb}$,
\begin{gather*}
\Sigma_{a_1\dots a_n}=\frac{1}{ (D-n)!}\epsilon_{a_1\dots a_D} e^{
a_{n+1}}\cdot\dots\cdot e^{a_D}
\end{gather*}
is a $(D-n)$-form constructed out of vielbeins $e^a$. The wedge
product between forms is supposed.

\subsection*{Acknowledgements}
Discussions with Igor Bandos, Martin Cederwall, Vladimir
Lyakhovsky, Dmitri Sorokin, Kellog Stelle, Mikhail Vasiliev,
Dmitri Vassilevich, Yuri Zinoviev are kindly acknowledged. Work
supported in part by the INTAS grant \#05-1000008-7928.

\pdfbookmark[1]{References}{ref}
\LastPageEnding

\end{document}